\documentclass[notoc]{JHEP3}
\preprint{MAN/HEP/2008/39\\CERN-TH/2008-211}

\usepackage{cite}
\usepackage{epsfig}
\usepackage{color}
\usepackage{rotating}
\usepackage{lscape}

\usepackage{graphics}
\usepackage{inputenc}
\usepackage{xspace}
\inputencoding{latin1}

\newcommand\justforus[1]{}

\newcommand{\T}{\ensuremath{\mathbf{T}}}

\newcommand{\M}{\ensuremath{\mathrm{M}}}
\newcommand{\N}{\ensuremath{\mathrm{N}}}
\renewcommand{\S}{\ensuremath{\mathrm{S}}}
\newcommand{\cK}{\ensuremath{\mathcal{K}}}
\newcommand{\cL}{\ensuremath{\mathcal{L}}}
\newcommand{\cM}{\ensuremath{\mathcal{M}}}
\newcommand{\cN}{\ensuremath{\mathcal{N}}}
\newcommand{\Gam}{\ensuremath{\mathbf{\Gamma}}}

\def\beq{\begin{equation}}
\def\eeq{\end{equation}}
\def\beqa{\begin{eqnarray}}
\def\eeqa{\end{eqnarray}}

\newcommand{\eqref}[1]{Eq.~(\ref{#1})\xspace}

\skip\footins = 1\bigskipamount plus 2pt minus 4pt                              
\title{\boldmath Symmetry of anomalous dimension matrices explained}

\author{
  Michael H. Seymour$^{ab}$ and Malin Sj\"odahl$^a$\\
  a School of Physics \& Astronomy, University of Manchester, \\
  Oxford Road, Manchester M13 9PL, U.K.\\
b Theoretical Physics Group, CERN, CH-1211 Geneva 23, Switzerland
}
  
  \abstract{
In a previous paper, one of us pointed out that the anomalous dimension
matrices for all physical processes that have been calculated to date
are complex symmetric, if stated in an orthonormal basis.  In
this paper we prove this fact and show that it is only true in a subset
of all possible orthonormal bases, but that this subset is the natural
one to use for physical calculations.
\justforus{Sections in oblique font are for our benefit and should not be
included in the final paper.}
  }

\begin{document}
 
 
\section{Introduction}
\label{sec:intro}

In perturbation theory resummation becomes necessary when large logarithms 
compensate the smallness of the coupling constant and invalidate a fixed 
order calculation.

For coloured processes the issue of resummation is complicated by the
non-Abelian colour structure.
For soft wide angle radiation, where real and virtual gluon emission cancel 
(at least for global observables), this complication can be expressed in 
terms of a matrix containing colour and phase space information, the 
soft anomalous dimension matrix
\cite{Sotiropoulos:1993rd,Contopanagos:1996nh,Kidonakis:1998nf, Oderda:1999kr,Bonciani:2003nt,Appleby:2003hp,Banfi:2004yd,Dokshitzer:2005ek,Kyrieleis:2005dt,Sjodahl:2008fz,Forshaw:2008cq},
\begin{equation}
\Gam=\sum_{i\not=j} \T_i\cdot\T_j \; \Omega_{ij},
\label{eq:Gamma}
\end{equation}
where $\T_i\cdot \T_j$ denotes the effect of exchanging a gluon between
partons $i$ and $j$ on the colour structure and $\Omega_{ij}$ is the result 
of integrating (azimuth and rapidity) over the region of phase space where 
emission is forbidden.  This integral can have an imaginary part, coming from
Coulomb phase contributions.

It has previously been observed that the soft anomalous 
dimension matrices have always been symmetric if stated in orthonormal
bases \cite{Seymour:2005ze}. 
Here we give the proof. The outline of the proof will be to first show 
that in orthonormal bases the colour matrix $\T_i\cdot \T_j$ is
Hermitian.  Then we show that for a particularly natural choice of
basis, it is also real.  From this the symmetry of $\T_i\cdot \T_j$, and
hence \Gam, follows.  We start by setting the notation.

\section{Notation}

We consider the amplitude, $\cM$, for a hard scattering process involving
$m$ partons (we do not need to make the distinction between incoming and
outgoing partons for the colour considerations discussed here).  In general
it is a function of the colour index of each of the $m$ partons,
\begin{equation}
  \cM = \cM_{a_1\ldots a_i\ldots a_j \ldots a_m},
\end{equation}
where $a_i$ is in the range 1~to~$N_c$ if $i$ is a quark or
antiquark and 1~to~$N_c^2\!-\!1$ if $i$ is a gluon.  In the present
discussion, it is only the colour of partons $i$ and $j$ that will be relevant,
so we suppress all other colour labels.  The set of physical colour
states forms a vector space, and it is useful to introduce a bra-ket
notation for this space,
\begin{equation}
  | M \rangle \equiv \cM_{\ldots a_i\ldots a_j\ldots}.
\end{equation}
Physical cross sections are given by the interference of amplitudes,
summed over colour indices,
\begin{equation}
  \sigma \sim \sum_{\ldots a_i\ldots a_j\ldots}
  \cN^*_{\ldots a_i\ldots a_j\ldots}\,\cM_{\ldots a_i\ldots a_j\ldots}
  \equiv \langle N | M \rangle.
\end{equation}
We are interested in the effect on the colour state of $\cM$ of
exchanging a gluon between partons $i$ and $j$, which can be written
\begin{equation}
  T_{a_ib_i}^\alpha\, T_{a_jb_j}^\alpha\, \cM_{\ldots b_i\ldots b_j\ldots},
\end{equation}
where $\alpha$ is the colour index of the exchanged gluon and $T$ is the
generator of the fundamental representation, $t_{a_ib_i}^\alpha$, if $i$
is an outgoing quark or incoming antiquark, its complex conjugate,
$(t_{a_ib_i}^\alpha)^* = t_{b_ia_i}^\alpha$, if $i$ is an outgoing
antiquark or incoming quark, and the generator of the
adjoint representation, $if_{a_i\alpha b_i}$, if $i$ is a
gluon\footnote{Different sign conventions are often chosen for physical
calculations, but that will not be relevant for the present
discussion.}.  Using the Hermiticty of the generators, the effect of
gluon exchange on the conjugate amplitude is
\begin{equation}
  (T_{a_ib_i}^\alpha)^* (T_{a_jb_j}^\alpha)^* \cM^*_{\ldots b_i\ldots b_j\ldots}=
  \cM^*_{\ldots b_i\ldots b_j\ldots} T_{b_ia_i}^\alpha T_{b_ja_j}^\alpha.
\end{equation}
Similarly, the interference of an amplitude formed by emitting a gluon
from parton $i$ in an amplitude $\cM$ with that formed by emitting a
gluon from parton $j$ in an amplitude $\cN$ is given by
\begin{equation}
  \cN^*_{\ldots b_i\ldots b_j\ldots} T_{b_ia_i}^\alpha T_{b_ja_j}^\alpha
  \cM_{\ldots a_i\ldots a_j\ldots}.
\end{equation}
It is therefore natural to define a colour operator $\T_i\cdot\T_j$, to
represent the exchange of a gluon between partons $i$ and $j$,
independent of where it lies relative to the cut through the diagram
that defines the physical final state,
\begin{equation}
  \label{eq:TiTj}
  \cN^*_{\ldots b_i\ldots b_j\ldots} T_{b_ia_i}^\alpha T_{b_ja_j}^\alpha
  \cM_{\ldots a_i\ldots a_j\ldots}
  \equiv \langle N | \T_i\cdot\T_j | M \rangle
  = \langle N | \T_j\cdot\T_i | M \rangle.
\end{equation}
The symmetry of the colour operator in its indices $i$ and $j$,
$\T_i\cdot\T_j=\T_j\cdot\T_i$ is obvious from its definition
(\ref{eq:TiTj})~-- since the two generators act on different partons'
indices their order is irrelevant.

In practical calculations it is convenient to introduce a basis for the
vector space of possible colour states for the hard process, the
elements of which we denote $|K\rangle$.  Any state $|M\rangle$ can be
written as a linear combination of $|K\rangle$ states,
\begin{equation}
  |M\rangle \equiv |K\rangle \M_K,
\end{equation}
where $\M_K$ are a set of complex numbers (represented as a column
vector \M).
\justforus{For now, we do not assume any special property of the basis (i.e.\
it is not orthogonal, normal or real) and define a matrix $\S$, called
the lowest order soft matrix by Sterman et al,
\begin{equation}
  \S_{KL} \equiv \langle K | L \rangle.
\end{equation}
It is evident from the definition that $\S$ is a Hermitian matrix.
The interference of two amplitudes $\cM$ and $\cN$ is given in the
matrix notation by
\begin{equation}
  \langle M | N \rangle = \M^\dagger \S \N.
\end{equation}
Eventually we will specialize to the case of an orthonormal basis, but}%
Restricting ourselves to orthonormal bases\footnote{Anomalous dimension
  calculations have typically been performed in orthogonal, but not
  normalized, bases, in which the symmetry property observed in
  \cite{Seymour:2005ze} is not manifest.  It is interesting to note that
  \cite{Nikolaev:2005zj}, which appeared on the same day as
  \cite{Seymour:2005ze}, does use an
  orthonormal basis and does obtain a symmetric anomalous dimension
  matrix, as does an appendix, not referred to in the rest of the paper,
  of \cite{Dokshitzer:2005ig}, which appeared some weeks later.}, we have
\justforus{%
\begin{equation}
  \langle K | L \rangle = \delta_{KL}
\end{equation}
and hence the interference of two amplitudes $\cM$ and $\cN$ is given in
the matrix notation by
\begin{equation}
  \langle M | N \rangle = \M^\dagger \N.
\end{equation}
From the interference of $\cM$ with a basis state $\cK$, we can obtain
an expression for the amplitude in the matrix representation,
which for the general basis is given by
\begin{equation}
  \M_K = (\S^{-1})_{KL} \, \langle L | M \rangle,
\end{equation}
and for the orthonormal basis by}
\begin{equation}
  \M_K = \langle K | M \rangle.
\end{equation}
Our aim is to find, and categorize the properties of, the matrix
representation of $\T_i\cdot\T_j$.  Acting with $\T_i\cdot\T_j$ on the
state $|M\rangle$ defines a new state, $\T_i\cdot\T_j|M\rangle$ from which we
obtain
\justforus{for the general basis
\begin{equation}
  \S_{LJ} (\T_i\cdot\T_j)_{JK} \M_K =
  \langle L | \T_i\cdot\T_j | K \rangle \M_K,
\end{equation}
and for the orthonormal basis,
\begin{equation}
  (\T_i\cdot\T_j)_{LK} \M_K =
  \langle L | \T_i\cdot\T_j | K \rangle \M_K.
\end{equation}
Since we want this to be true for any state $\M$, we must have
in the general basis
\begin{equation}
  (\T_i\cdot\T_j)_{JK} =
  (\S^{-1})_{JL}\langle L | \T_i\cdot\T_j | K \rangle,
\end{equation}
and in the orthonormal basis,}
\begin{equation}
  \label{eq:matrixTiTj}
  (\T_i\cdot\T_j)_{LK} =
  \langle L | \T_i\cdot\T_j | K \rangle.
\end{equation}
We can now state the goal of our proof.  Since we observe that the
matrix representation of $\Gamma$ is symmetric for arbitrary
observables, and the $\Omega_{ij}$ are observable-dependent, it must be
that the matrix representation of $\T_i\cdot\T_j$ is symmetric in
orthonormal bases.  Our proof is in two parts, first we prove that it
is Hermitian in any orthonormal basis, then we try to prove that it
is real, since a real Hermitian matrix is automatically symmetric.
However, we will find that it is true in only a subset of bases.
Finally, we will argue that these are the most natural set of bases, and
hence that it is not surprising that all previous calculations, when
orthonormalized, do give symmetric matrices.

\section{Hermiticity of colour structure}

Having taken care to set up the notation, the proof that the matrix
representation of $\T_i\cdot\T_j$ is Hermitian
\justforus{in an orthonormal basis}%
is straightforward.  It is a consequence only of the definition
(\ref{eq:matrixTiTj}) and the Hermiticity of the generators in the
parton indices.
In fact it is already clear from the definition of the matrix element
(\ref{eq:TiTj}) that the right-hand-side of Eq.~(\ref{eq:matrixTiTj}) is
Hermitian,
\begin{eqnarray}
  ((\T_i\cdot\T_j)_{LK})^* =
  (\langle L | \T_i\cdot\T_j | K \rangle)^* &=&
  (\cL^*_{\ldots b_i\ldots b_j\ldots} T_{b_ia_i}^\alpha T_{b_ja_j}^\alpha
  \cK_{\ldots a_i\ldots a_j\ldots})^*
\nonumber\\
  &=& \cK^*_{\ldots a_i\ldots a_j\ldots} T_{a_ib_i}^\alpha T_{a_jb_j}^\alpha
  \cL_{\ldots b_i\ldots b_j\ldots}
\nonumber\\
  &=& \langle K | \T_i\cdot\T_j | L \rangle =
  (\T_i\cdot\T_j)_{KL}.
\end{eqnarray}
That is,
\begin{equation}
  (\T_i\cdot\T_j)^\dagger = \T_i\cdot\T_j,
\end{equation}
the matrix representation of $\T_i\cdot\T_j$ is Hermitian.

\justforus{In a general basis, on the other hand, we have
\begin{eqnarray}
  ((\T_i\cdot\T_j)^\dagger)_{KJ} &=&
  ((\T_i\cdot\T_j)_{JK})^* =
  [(\S^{-1})_{JL}\langle L | \T_i\cdot\T_j | K \rangle]^*
  \nonumber\\
  &=& [(\S^{-1})_{JL}]^*[\langle L | \T_i\cdot\T_j | K \rangle]^*
  = \langle K | \T_i\cdot\T_j | L \rangle(\S^{-1})_{LJ},
\end{eqnarray}
where we have used the fact that $\S$ is Hermitian.  However, in general
$\S$ will not commute with $\langle \T_i\cdot\T_j \rangle$, so
\begin{eqnarray}
  (\T_i\cdot\T_j)^\dagger \not= \T_i\cdot\T_j.
\end{eqnarray}
Even restricting to an orthogonal basis, which has
$\S_{KL}=N_{(K)}\delta_{KL}$ with $N_{(K)}$ real, we have
\begin{eqnarray}
  ((\T_i\cdot\T_j)^\dagger)_{KL} &=&
  ((\T_i\cdot\T_j)_{LK})^* =
  [\frac1{N_{(L)}}\langle L | \T_i\cdot\T_j | K \rangle]^*
  = \frac1{N_{(L)}}\langle K | \T_i\cdot\T_j | L \rangle
  \nonumber\\
  &\not=& 
  (\T_i\cdot\T_j)_{KL}
  = \frac1{N_{(K)}}\langle K | \T_i\cdot\T_j | L \rangle,
\end{eqnarray}
unless all the $N_{(K)}$ are equal, i.e.\ the basis is normalized.
}

\section{Realness of scalar product}

It now remains only to show that $\langle K|\T_i\cdot\T_j|L\rangle$ is
real for all $i$ and $j$ for all elements $|K\rangle$, $|L\rangle$ in an
orthonormal basis.  In fact it is easy to see that this cannot be the
case in general, because if there is some initial basis in which it is
true, one can apply an arbitrary unitary transformation to another
orthonormal basis in which it is not.  Our task is therefore to show
that there does exist a basis, or set of bases, in which it is true.  In
this section we show, not only that this is the case, but also that such
a basis is the most natural one to use for a physical calculation,
explaining why this property has been observed in all previous
calculations.

In general, one can always choose a basis constructed from delta
functions in quark indices, delta functions in gluon indices, generators
$t^a_{bc}$, and the symmetric and antisymmetric structure constants
$d_{abc}$ and $i f_{abc}$,
\begin{equation}
(if/d)_{abc}=2(\mbox{Tr}[t^a t^b t^c] (-/+) \mbox{Tr}[t^b t^a t^c]).
\label{eq:fd}
\end{equation}
These can be used to form a complete basis for any physical process,
because the Feynman rules contain no other factors.  Therefore the
matrix element $\langle K|\T_i\cdot\T_j|L\rangle$ is a scalar in colour
space, constructed from these building blocks.  Using the relation
(\ref{eq:fd}), all occurences of $d_{abc}$ and $i f_{abc}$ can be
converted into generators of the fundamental representation.  Moreover,
these generators can be replaced pairwise by delta functions using the
relation
\begin{equation}
t^\alpha_{ca} t^\alpha_{db}
=\frac{1}{2}(\delta_{ad}\delta_{bc}-\frac{1}{N_c}\delta_{ac}\delta_{bd}).
\label{eq:tata}
\end{equation}
Thus every matrix element in such a basis reduces to strings of delta
functions with real coefficients and is clearly real.

Therefore any orthonormal basis constructed from delta functions,
generators and group structure constants will result in a symmetric
anomalous dimension matrix.  As we have already mentioned, such a choice
of basis is extremely natural for describing physical processes, since
the colour part of any Feynman rule can be represented in this way.
Although we have shown that the colour part of the anomalous dimension
matrix, $\T_i\cdot\T_j$ in Eq.~(\ref{eq:Gamma}), is real and Hermitian,
and therefore symmetric, the kinematic part, $\Omega_{ij}$, is complex
in general, so the anomalous dimension matrix itself is complex
symmetric, and hence not Hermitian.

\section*{Acknowledgments}
We thank Jeff Forshaw, Johan Gr\"onqvist and G\"osta Gustafson for
useful discussions. MS thanks the Theoretical Physics Department of
Lund University for their hospitality during the completion of this
work.

\bibliographystyle{utcaps}  
\bibliography{RRefs} 

\end{document}